\documentclass{article}
\usepackage{PRIMEarxiv}
\usepackage[T1]{fontenc}
\usepackage[utf8]{inputenc}
\usepackage{microtype}
\usepackage{amsmath,amssymb}
\usepackage{graphicx}
\usepackage{xcolor}
\usepackage[numbers,sort&compress]{natbib}
\usepackage{url}
\usepackage[hidelinks]{hyperref}
\hypersetup{hypertexnames=false}
\usepackage{booktabs}
\usepackage{placeins}
\usepackage{pdfpages}
\title{Differentiable powder X-ray diffraction enables analysis of evolving structures}
\author{
  Xiangyu Yin$^{*}$, Nina Andrejevic, Hemant Sharma, Ming Du, Aileen Luo, Michael Prince, \\ 
  \textbf{Burak Guzelturk, Xiaoyi Zhang, Brian H. Toby, Mathew J. Cherukara$^{\dagger}$} \\
  Advanced Photon Source, Argonne National Laboratory, Lemont IL, USA \\
  * xyin@anl.gov \\
  $\dagger$ mcherukara@anl.gov
}

\begin{document}
\pagenumbering{gobble}
\thispagestyle{empty}
\textbf{GOVERNMENT LICENSE}

The submitted manuscript has been created by UChicago Argonne, LLC, Operator of Argonne National Laboratory (“Argonne”). Argonne, a U.S. Department of Energy Office of Science laboratory, is operated under Contract No. DE-AC02-06CH11357. The U.S. Government retains for itself, and others acting on its behalf, a paid-up nonexclusive, irrevocable worldwide license in said article to reproduce, prepare derivative works, distribute copies to the public, and perform publicly and display publicly, by or on behalf of the Government. The Department of Energy will provide public access to these results of federally sponsored research in accordance with the DOE Public Access Plan. \href{http://energy.gov/downloads/doe-public-access-plan}{http://energy.gov/downloads/doe-public-access-plan}
\clearpage
\pagenumbering{arabic}
\raggedbottom
\maketitle

\begin{abstract}
\emph{Operando} and \emph{in-situ} experiments extend powder X-ray diffraction (PXRD) to materials that evolve during experiments, producing series of hundreds or thousands of patterns. Such series are commonly refined one pattern at a time, so weak signals give noisy structural trajectories. Parametric refinement instead fits a shared model to all patterns, but with most existing programs, changing the model or fitting it efficiently remains challenging. Here, we introduce diffPXRD, a differentiable PXRD engine in PyTorch. diffPXRD calculates derivatives by automatic differentiation, so a shared model can be changed without new derivative code. Batched fitting is also more than seven times faster than serial fitting. In an \emph{operando} cathode series, a shared lattice model can reduce lattice-parameter variability by 68.4\% when 10\% of the photon counts are retained. We also show shared models can describe the precipitation and dissolution of Li$_2$S and reconstruct collective atomic displacements in Ca$_3$Ti$_2$O$_7$.
\end{abstract}


\section{Introduction}

Powder X-ray diffraction (PXRD) is a widely used technique for determining crystal structures, with \emph{operando}, \emph{in-situ}, and time-resolved pump-probe measurements extending its role to the study of evolving materials \cite{mikheenkova2023resolving,chien2023rapid,pomiro2020entropy,herve2024ultrafast}. Changes in peak positions reveal lattice expansion or contraction, while reflection intensities provide information about atomic arrangements and crystalline phase fractions, and peak widths and shapes about crystallite size and microstrain. Quantitative interpretation requires connecting these observations to a structural model through the diffraction calculation. In practice, PXRD series are commonly refined one pattern at a time, so weak signals give noisy trajectories.

Established refinement methods already show ways to connect PXRD patterns to structures. Rietveld refinement adjusts a crystal model to reproduce the full measured profile \cite{rietveld1969profile}, whereas Pawley and Le Bail methods refine lattice parameters and reflection intensities without determining atomic coordinates \cite{pawley1981unit,lebail1988abinitio}. Applied sequentially, as in GSAS-II, these methods can determine structural evolution by refining successive measurements \cite{toby2013gsasii}. Parametric refinement takes a different approach by linking parameters across the series through shared parametric functions \cite{stinton2007parametric}. For a diffraction series, such as one recorded during battery cycling or heating, a parametric function can describe how selected structural parameters depend on the experimental conditions. The parameters of these functions are fitted directly to the profiles, making the relationship between experimental conditions and structure part of the refinement itself. TOPAS handles such relationships through user-defined equations and computer algebra, determining parameter dependencies for derivatives \cite{coelho2018topas}. 

Recent studies have used differentiable diffraction calculation for structure optimization against simulated powder patterns \cite{segal2026loss} and for electron diffraction refinement with learned experimental corrections \cite{malik2026hybrid}. Differentiable diffraction calculation enables automatic differentiation (AD), which follows the numerical operations in the diffraction computational graph and applies the chain rule to obtain its derivatives with respect to all variables. This allows a diffraction calculation to be composed with a shared parametric model expressed as tensor operations, so that both are differentiated and executed in the same computational process. This provides a programmable implementation in which researchers can change the shared model without changing the diffraction engine or implementing its derivatives separately. For example, a phase fraction can follow either a flexible spline or a kinetic law, and automatic differentiation supplies the derivatives of both.

Although simple implementations of differentiable diffraction exist \cite{segal2026loss,malik2026hybrid}, we note that applying this approach to experimentally measured PXRD requires a forward calculation that also reproduces instrumental broadening and sample effects. Experimental diffraction series can contain hundreds or thousands of patterns, making efficient evaluation of profiles and derivatives essential for refinement. Previous studies have accelerated powder diffraction structure determination with graphics processing units (GPUs) \cite{spillman2021gallop} and pattern generation with optimized CPU implementations \cite{lebeda2026xrdrust}. Our goal is to combine this hardware acceleration with a differentiable forward model that includes these effects, making it easier to develop and compare structural and shared parametric models against measured diffraction series.

Here, we present diffPXRD, a differentiable constant-wavelength PXRD engine that combines these capabilities within PyTorch \cite{paszke2019pytorch}. Our implementation supports independent structural refinement and joint analysis through shared parametric models, with automatic gradients and batched CPU or GPU evaluation. We first validate the diffraction calculation against GSAS-II, check its derivatives, and characterize its computational performance. Then we present three applications that connect empirical and physically constrained models to measured series: cathode lattice evolution, where an empirical curve stabilizes weak signals, precipitation and dissolution during battery cycling, where a shared kinetic shape transfers to later cycles and predicts their unfitted profiles, and collective atomic displacements on heating and cooling, where a neutron-derived basis limits the allowed motions.

\section{Results}

\subsection{Differentiable diffraction and series refinement}

\begin{figure}[!ht]
\centering
\includegraphics[width=\linewidth]{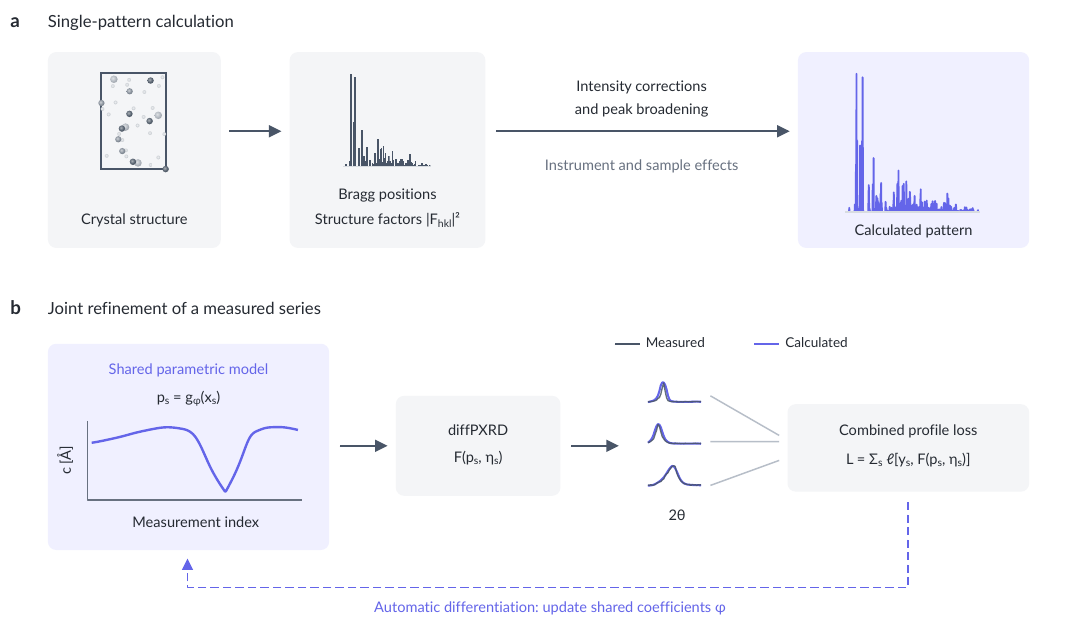}
\caption{\textbf{Differentiable diffraction and joint series refinement.} \textbf{a}, The crystal structure determines reflection positions and structure factors. Instrumental and sample corrections determine the reflection intensities and peak shapes that form the calculated powder pattern. The crystal and pattern illustrate a single calculation. \textbf{b}, A shared parametric model supplies the structural parameters $\mathbf{p}_s$ at each experimental state $x_s$, while parameters fitted separately for each pattern, such as peak widths, are collected in $\boldsymbol{\eta}_s$. diffPXRD calculates the corresponding patterns, and their differences from the measured profiles contribute to a combined loss. Automatic differentiation gives the gradient used to update the shared coefficients $\phi$. The example trajectory and measured profiles are from the nickel-cobalt-aluminum oxide (NCA) cathode series in Fig.~\ref{fig:operando}. Calculated profile overlays are schematic.}
\label{fig:overview}
\end{figure}

The diffPXRD engine calculates a powder pattern from a structural model through differentiable PyTorch operations, allowing the same implementation to generate both profiles and the gradients used for refinement. Figure~\ref{fig:overview} illustrates this connection for a series of measurements. Each experimental state supplies a set of structural parameters to the diffraction calculation, which determines Bragg positions and structure factors, applies intensity corrections, and broadens the reflections onto the measurement grid. The reflection list is fixed, but reflection positions and structure factors vary continuously with the lattice parameters and atomic coordinates, so every step can be differentiated. Comparing the calculated and measured profiles defines the fitting error, or loss. Its gradient describes how changes in the lattice, atomic, instrument, and sample parameters affect agreement with the data and therefore supplies the direction for an optimization step (Methods).

For diffraction series, the structural parameters can be fitted independently to each pattern, which we call \emph{framewise refinement}, or linked through a shared parametric model for \emph{joint refinement}. Following parametric refinement \cite{stinton2007parametric}, the shared model expresses selected structural parameters as functions of the experimental conditions, $\mathbf{p}_s=g_\phi(x_s)$. Here $x_s$ denotes the temperature, time, or capacity at measurement $s$, together with the charge/discharge or heating/cooling direction when the path depends on it, and $\phi$ contains the coefficients shared by all patterns. We refer to the path that $\mathbf{p}_s$ follows through the series, such as the variation of a cathode lattice parameter with capacity, as the structural trajectory. Parameters that are not linked across the series, such as peak widths, are still fitted to each pattern and collected in $\boldsymbol{\eta}_s$. With the diffraction calculation $F$, the total loss relative to the measured patterns $y_s(2\theta)$ is
\begin{equation}
\mathcal{L}(\phi,\{\boldsymbol{\eta}_s\})=\sum_s\ell\left[y_s,F\left(g_\phi(x_s),\boldsymbol{\eta}_s\right)\right],
\label{eq:joint}
\end{equation}
where $\ell$ measures the disagreement for an individual pattern. A shared coefficient can influence several calculated patterns, and differentiation through $g_\phi$ and $F$ combines their contributions to determine its gradient (Fig.~\ref{fig:overview}b). The coefficients are thus fitted against the profiles themselves rather than only against parameters extracted from separate refinements. Replacing the shared model changes $g_\phi$, while the diffraction calculation and automatic evaluation of its derivatives remain the same.

diffPXRD evaluates multiple patterns together in batches on CPUs or GPUs. We note that batching and parameter sharing serve different purposes. Batching reduces the overhead of separate calculations without imposing relationships between the fitted structures, whereas parameter sharing changes which structural trajectories the model can represent. Both framewise and joint refinement can therefore be batched, and we compare these arrangements separately in the following examples. Details of the diffraction calculation and the treatment of scale and background are given in Methods.

\FloatBarrier
\subsection{Numerical accuracy and computational performance}

\begin{figure}[!ht]
\centering
\includegraphics[width=0.94\linewidth]{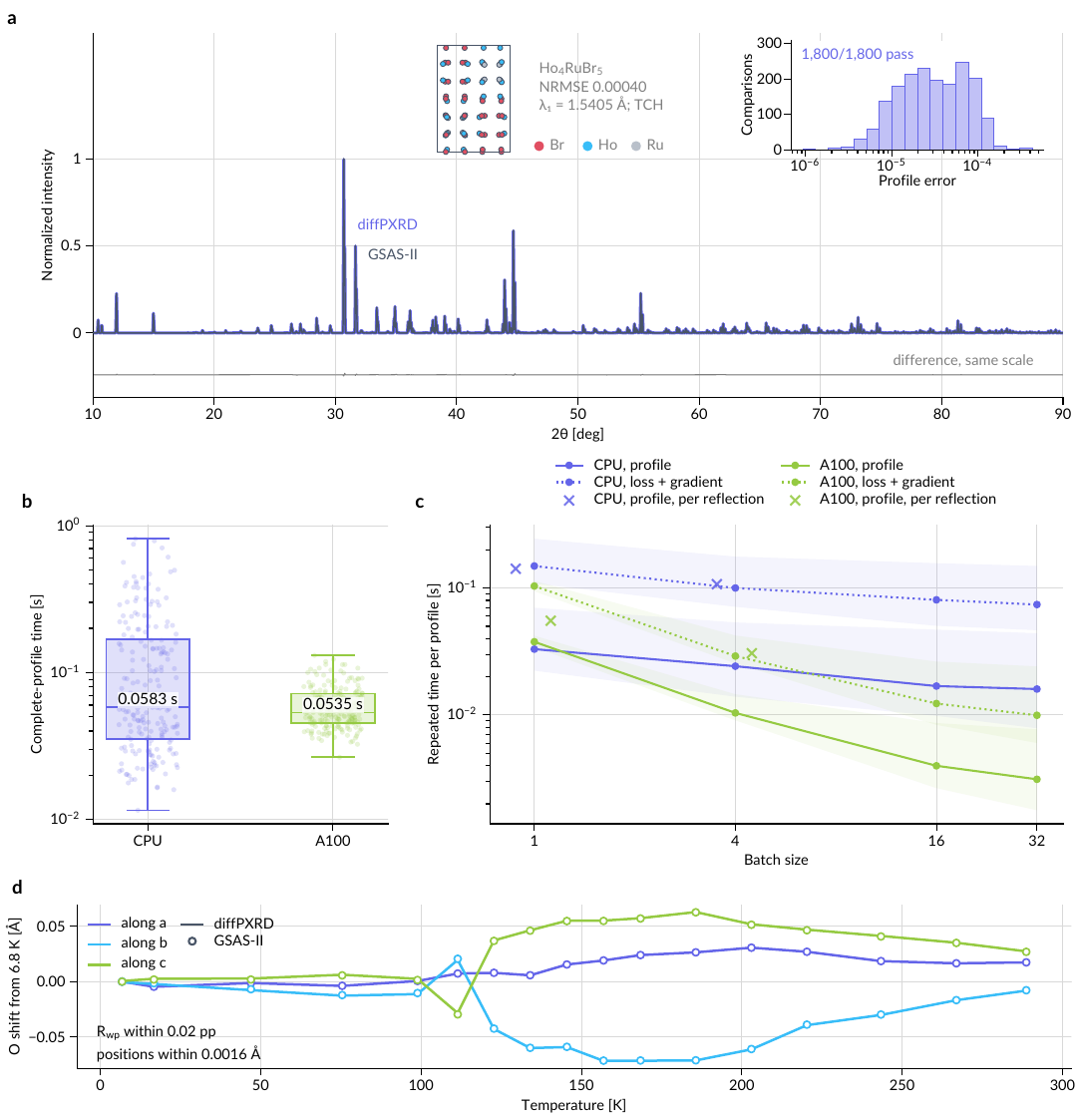}
\caption{\textbf{Numerical agreement and computational performance.} \textbf{a}, The comparison with the largest profile error among the 1,800 tests against GSAS-II, for tetragonal Ho$_4$RuBr$_5$ with Cu~K$\alpha$ radiation and a Thompson-Cox-Hastings (TCH) peak shape. The difference curve is offset below the two profiles on the same intensity scale. The histogram shows the normalized root-mean-square error (NRMSE) of all 1,800 tests. NRMSE is the RMS profile difference divided by the maximum of the reference profile, after fitting an overall intensity scale. \textbf{b}, Time to calculate one complete pattern for each of the 200 ICSD structures, including reading the structure file and building the reflection list. Each point is one structure. Boxes span the interquartile range and whiskers the full range. \textbf{c}, Time per profile when a prebuilt reflection list is evaluated repeatedly, as in refinement. Solid lines show the profile alone and dotted lines a least-squares loss with its gradient with respect to two lattice-scale parameters per pattern. Lines are medians and bands are interquartile ranges over the same 200 structures. Crosses, offset for clarity, show the profile time at batch sizes 1 and 4 when every reflection is broadened separately. Elsewhere in \textbf{b} and \textbf{c}, symmetry-equivalent reflections are merged into one peak. All timings use FCJ profiles in double precision on four CPU threads or one A100 GPU (Methods). \textbf{d}, Shift of the refined oxygen position from its 6.8~K value along each cell axis for the 17 measured CuCr$_2$O$_4$ patterns, from diffPXRD (lines) and GSAS-II (open circles).}
\label{fig:performance}
\end{figure}

We first tested whether diffPXRD reproduces the profiles calculated by GSAS-II. To this end, we selected 200 structures from the Inorganic Crystal Structure Database (ICSD) \cite{belsky2002icsd}. The structures span all seven crystal systems and a wide range of space groups and unit cell sizes. We calculated each structure with both programs under nine instrument and sample configurations, from laboratory to synchrotron sources (Methods). All 1,800 comparisons pass our predefined acceptance criteria for reflection lists, positions, intensities, and profiles (Supplementary Note~1). Figure~\ref{fig:performance}a shows the comparison with the largest normalized root-mean-square (RMS) error. Even for this structure, the reflection positions and integrated intensities agree closely. The derivatives calculated by AD also agree with central finite differences (Supplementary Note~1). The agreement in both profiles and derivatives supports the use of diffPXRD for refinement.

We then compared the speed of diffPXRD on a CPU and a GPU, using profiles with the Finger-Cox-Jephcoat (FCJ) axial-divergence correction \cite{finger1994correction}. For a complete calculation of a single pattern, the median times on the two devices are comparable (Fig.~\ref{fig:performance}b). However, the CPU time grows with the number of peaks, whereas the GPU time is dominated by fixed costs and changes little. The CPU is therefore faster for most small structures, and the GPU is faster for every structure with more than 500 peaks. For the largest fifth of the structures, the median GPU speedup is nearly fourfold (Supplementary Note~2).

Refinement uses the calculation differently. The reflection list is built once, and the profile is then recalculated many times as the parameters change, often for many patterns at once. We timed these repeated evaluations for the profile alone and for a least-squares loss with its gradient (Fig.~\ref{fig:performance}c). Reusing the reflection list removes the setup cost on both devices, but for a single pattern the two remain comparable. As more patterns are evaluated together, the time per profile drops on both devices, and much more on the GPU. At 32 patterns per batch, the median GPU speedup reaches 7.4 for the loss and gradient that refinement requires.

These timings merge symmetry-equivalent reflections into one peak, as in GSAS-II. Merging is exact only while the lattice keeps its symmetry. A refinement that lowers the symmetry, for example from cubic to tetragonal, must instead regroup the split reflections or broaden them separately. With every reflection broadened separately, the GPU is faster even for a single pattern (crosses in Fig.~\ref{fig:performance}c). The GPU advantage in refinement therefore comes mainly from batching and from larger numbers of peaks.

Finally, we compared the two programs on the full-profile refinement of measured data, using the variable-temperature CuCr$_2$O$_4$ series distributed with GSAS-II (Fig.~\ref{fig:performance}d). In each pattern, both programs refine the three oxygen coordinates, two phase amplitudes, and the background. The lattice, broadening, and displacement parameters are fixed at the same calibrated values. No shared temperature dependence is imposed. The refined oxygen positions from the two programs agree to within 0.0016~\AA{} across the series, including the abrupt shift near the tetragonal-to-orthorhombic transition at about 130~K \cite{suchomel2012spin}, and their weighted profile R-factors ($R_{\mathrm{wp}}$) differ by at most 0.02 percentage points (Supplementary Note~1). The two programs also respond similarly to the starting model. When started from oxygen positions displaced by up to 0.3~\AA{}, both reach the same refined positions in nearly all refinements (Supplementary Note~1).

\FloatBarrier
\subsection{\emph{Operando} lattice evolution}

\begin{figure}[!ht]
\centering
\includegraphics[width=\linewidth]{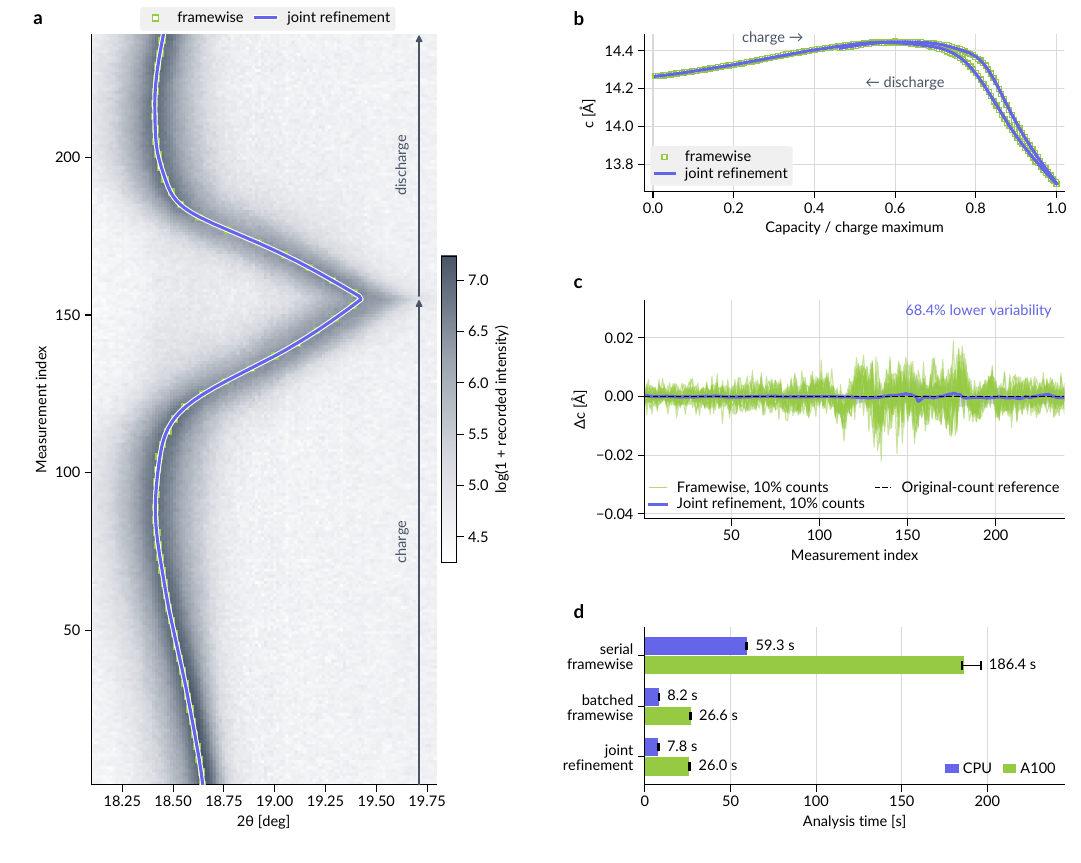}
\caption{\textbf{Lattice evolution and noise stability in the NCA series.} \textbf{a}, Measured intensity near the 003 reflection during charge and discharge. Overlays show peak positions from framewise refinement at every fourth pattern and from the shared capacity model with 12 spline coefficients per branch. \textbf{b}, Refined $c$ as a function of capacity. \textbf{c}, Changes in $c$ relative to the joint refinement of the original-count data, for twenty paired realizations retaining 10\% of the counts on average. Thin green lines show the individual framewise trajectories, and the purple line and band show the median and full range of joint trajectories. The dashed zero line marks the original-count reference. \textbf{d}, Analysis times for serial framewise, batched framewise, and joint refinement on four CPU threads and one A100, including input reading and output writing. Bars show medians and whiskers span three repetitions.}
\label{fig:operando}
\end{figure}

To demonstrate the analysis of evolving structures, we applied diffPXRD to published \emph{operando} measurements of a nickel-cobalt-aluminum oxide (NCA) cathode during charge and discharge \cite{mikheenkova2023resolving,mikheenkova2022resolvingdata}. The position of the 003 reflection follows the layer spacing $d_{003}=c/3$, providing a direct measure of lattice evolution. Its movement to lower angle during the initial part of charge indicates expansion along $c$, followed by contraction as the reflection moves to higher angle near the end of charge (Fig.~\ref{fig:operando}a). Discharge largely reverses this evolution, with different charge and discharge paths at the same capacity (Fig.~\ref{fig:operando}b).

We described this behavior with empirical cubic B-splines, which join local polynomial segments into a smooth dependence of $c$ on capacity (Eq.~\ref{eq:nca_shared_model}). Charge and discharge have separate curves, allowing their shapes to differ rather than constraining their separation to an offset or slope. Twelve coefficients per branch replace an independently fitted lattice parameter for each of the 239 patterns. We chose this number as the smallest that matched framewise refinement within predefined limits on the training profiles, before any omitted patterns were evaluated (Methods). The Gaussian and Lorentzian widths of the pseudo-Voigt peak shape are fitted separately for each pattern in both methods, so that changes in peak breadth do not have to be absorbed into the lattice trajectory.

At the original intensity level, the shared curve and framewise estimates give similar lattice trajectories (Fig.~\ref{fig:operando}b). Parameter sharing can stabilize the trajectory when the signal is weaker, because each coefficient is then constrained by several patterns rather than a single measurement. To test this, we generated twenty noisy versions of the series, each retaining on average 10\% of the measured counts, and refined every version with both methods (Methods). The framewise trajectories fluctuate between neighboring patterns, whereas the joint trajectories retain the expansion and contraction with a narrower spread (Fig.~\ref{fig:operando}c), reducing the RMS repeat-to-repeat standard deviation of $c$ from 0.00340 to 0.00107~\AA{}, a reduction of 68.4\%, while preserving agreement with the original profiles (Supplementary Note~3). The lower variability comes from the smoothness that the shared curve imposes, which a curve fitted afterwards to the framewise values would also impose. It reflects that assumption rather than a smaller measurement uncertainty. Fitting the curve to the profiles instead lets the measured patterns, rather than one extracted value per pattern, determine where the trajectory bends, and the omitted patterns test that choice against profiles not used in the fit (Supplementary Note~3).

Joint and batched framewise fitting have similar analysis times, both substantially shorter than serial framewise fitting (Fig.~\ref{fig:operando}d). On four CPU threads, the median times are 7.85, 8.22, and 59.33~s, respectively. The small 003-window workload is faster on the CPU than on the A100, where joint fitting takes 25.98~s. This is consistent with the fixed GPU costs that dominate small calculations in Fig.~\ref{fig:performance}b. Thus, the main timing benefit in this application comes from evaluating the series in batch, while the shared model provides the more stable lattice trajectory.

\FloatBarrier
\subsection{\texorpdfstring{Li$_2$S}{Li2S} precipitation and dissolution}

\begin{figure}[!ht]
\centering
\includegraphics[width=\linewidth]{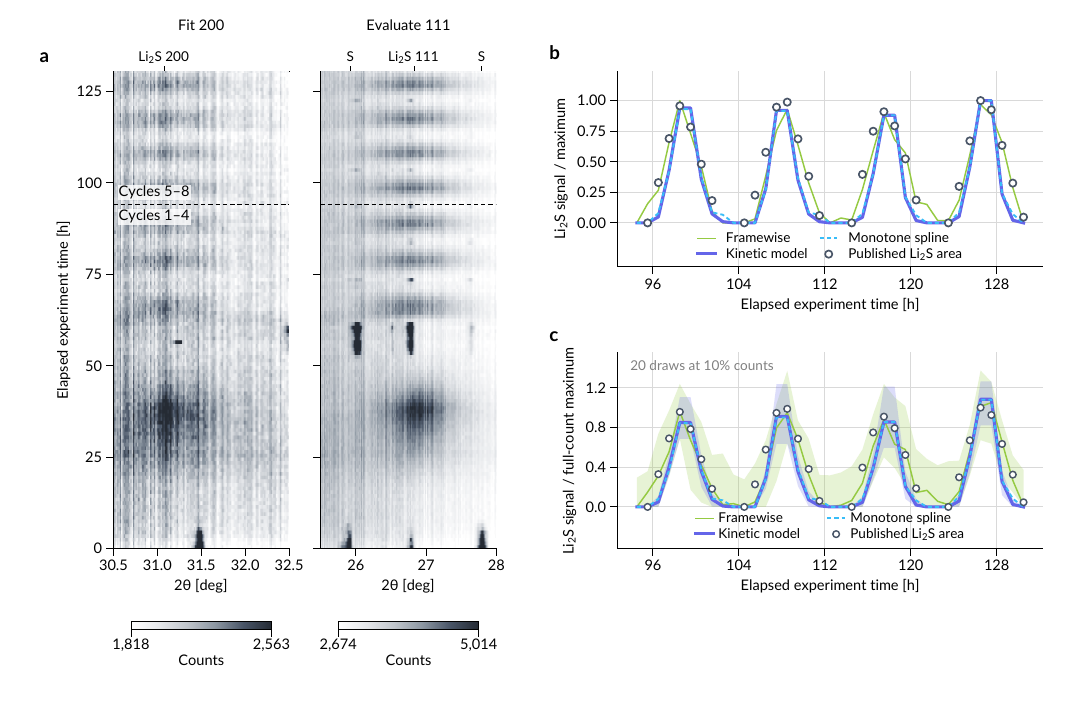}
\caption{\textbf{Precipitation and dissolution of crystalline Li$_2$S.} \textbf{a}, Measured intensity in the 200 fitting window and the 111 evaluation window over 131 patterns. Each map uses its own linear count scale. Dashed lines separate cycles 1--4, used to fit the shared curve shape, from evaluation cycles 5--8. Labels mark the broad Li$_2$S 200 and 111 peaks and two sharp sulfur (S) peaks. Sharp peaks from other phases are not part of the Li$_2$S model (Supplementary Note~3). \textbf{b}, Predicted Li$_2$S intensity trajectories in the evaluation cycles at the original count level, normalized to their maxima for comparison with published 111 integrated areas. \textbf{c}, Median trajectories over twenty paired realizations at 10\% count retention. Bands show central 95\% ranges for framewise and kinetic fits, scaled by each model's fixed original-count maximum. Published 111 areas provide the reference in both \textbf{b} and \textbf{c}.}
\label{fig:lisulfur}
\end{figure}

Time series analysis can also follow phase formation and dissolution through changes in a crystalline phase's contribution to the pattern. In the published lithium-sulfur series of Chien et al. \cite{chien2020simultaneous,chien2020simultaneousdata}, Li$_2$S reflections emerge during discharge and disappear during charge. Figure~\ref{fig:lisulfur}a shows the weak 200 and 111 reflections across the series. The 200 window shows the repeated appearance of Li$_2$S, motivating a description that shares the precipitation and dissolution behavior across cycles rather than fitting each intensity independently.

We used a kinetic model with four shared shape parameters for delayed precipitation and subsequent dissolution \cite{fan2015mechanism}. An onset and a growth exponent describe the increase in Li$_2$S signal during discharge, while an exponential decay with two shape parameters describes its disappearance during charge. Both branches follow normalized capacity within each half-cycle. A cycle-dependent amplitude accommodates changes in signal magnitude without changing the shared shape. This amplitude represents relative crystalline Li$_2$S intensity rather than an absolute phase fraction.

\begin{figure}[!t]
\centering
\includegraphics[width=\linewidth]{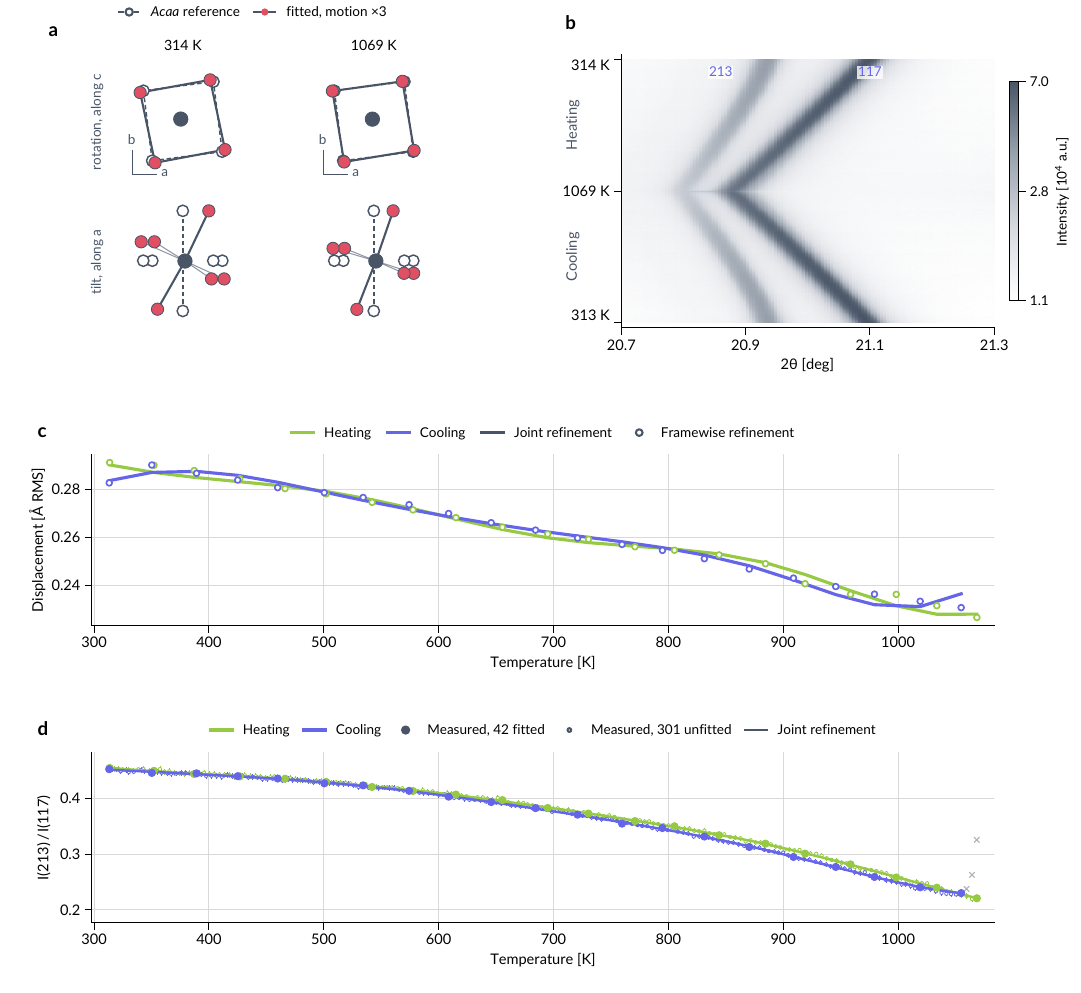}
\caption{\textbf{Collective displacements in Ca$_3$Ti$_2$O$_7$.} \textbf{a}, A TiO$_6$ octahedron at 314 and 1069~K on heating, viewed along $c$ (top) and an in-plane axis (bottom). Open sites mark the $Acaa$ reference. Displacements are amplified threefold in a fixed cell, with oxygen in red and titanium in gray. \textbf{b}, All 346 measured 213/117 doublet profiles during heating and cooling, displayed on a logarithmic intensity scale after averaging in 0.005$^\circ$ bins. \textbf{c}, RMS displacement of the 48 atomic mean positions from $Acaa$ as a function of temperature, from joint (lines) and independently initialized framewise (open circles) refinement, with heating in green and cooling in purple. The two thermal branches are fitted without a continuity constraint at the turnaround. \textbf{d}, Measured $I(213)/I(117)$ at the 42 fitted (filled) and 301 unfitted (open) temperatures, compared with ratios calculated from the fitted atomic splines (lines). Error bars are uncertainties of the two-peak intensity ratios, not of the atomic coordinates. Gray crosses mark three cooling patterns just below the phase transition, which are excluded because the 213 peak is still shifted (Supplementary Note~4).}
\label{fig:atomic-trajectory}
\end{figure}

We evaluated whether the shared shape transfers to later cycles and another reflection. A fixed diffraction template, calibrated using the 200 and 111 reflections in cycles 1-4, relates their positions, widths, and relative intensities. Only the 200 reflection in these cycles is used to fit the shared trajectory shape. In cycles 5-8, the template and shape remain fixed, while cycle amplitudes and per-pattern backgrounds are fitted to 200. The resulting intensity trajectory then predicts the 111 signal without fitting that reflection in the evaluation cycles. We compare these predictions with published 111 integrated areas and with a more flexible monotone spline and independent framewise amplitudes evaluated under the same protocol (Methods).

Both shared models reproduce the repeated rise and decay of the published integrated areas in the evaluation cycles (Fig.~\ref{fig:lisulfur}b). At the original counts, the framewise amplitudes follow the published areas most closely, and the shared and framewise trajectories differ most where the Li$_2$S signal is weak or vanishing. At 10\% intensity retention, the kinetic and spline trajectories remain similar, while the framewise trajectories develop a wider range between noise realizations (Fig.~\ref{fig:lisulfur}c). The shared models then correlate better with the published areas than framewise fitting does, with a median Pearson correlation of 0.90 compared with 0.83 over the twenty realizations, while their scaled RMS differences from those areas remain similar, 0.36 for the kinetic model and 0.34 for framewise fitting (Supplementary Note~3). The kinetic model reduces trajectory variability by about half relative to framewise refinement, with a modest increase in the unfitted 111 profile loss (Supplementary Note~3). Sharing the curve shape across cycles thus trades a lower correlation with the published areas at the original counts for reduced variability at low counts. The shared shape parameters describe the evolution of the Li$_2$S signal without uniquely identifying a microscopic precipitation mechanism.

\FloatBarrier
\subsection{Atomic motions during thermal treatments}

Relative reflection intensities provide sensitivity to atomic motions within a phase, extending time series analysis beyond lattice parameters and crystalline phase fractions. We demonstrate this with a published Ca$_3$Ti$_2$O$_7$ series, whose polar $A2_1am$ structure involves coupled octahedral rotations and tilts and evolves toward the nonpolar $Acaa$ structure on heating \cite{pomiro2020entropy,pomiro2020data}. Rather than refine every atomic coordinate independently, we describe the structure with three collective displacement vectors derived from neutron refinements of a separate sample \cite{kong2023scaling}. The neutron structures specify allowed coordinated movements that combine octahedral rotation, tilt, and cation displacement. The X-ray profiles can then determine their amplitudes, with temperature-dependent splines describing the heating and cooling paths. We note that the trajectories in this example describe changes in atomic mean positions, rather than tracking the dynamics of individual atoms.

Figure~\ref{fig:atomic-trajectory}a illustrates the resulting motion of a TiO$_6$ octahedron relative to the $Acaa$ reference, with reduced tilt on heating visible in the in-plane projection. The neighboring 213 and 117 reflections change both separation and relative intensity through the thermal cycle (Fig.~\ref{fig:atomic-trajectory}b). Their separation tracks lattice changes, while their relative intensities provide sensitivity to the collective atomic displacements. We fit 42 approximately evenly spaced temperatures, covering both branches while leaving the intervening patterns for evaluation. Ten reflection windows constrain the profiles, with an additional weight on the integrated 213/117 ratio from the same patterns. This gives the relative peak areas explicit influence when the effective profile model does not reproduce every detail of the measured line shape (Methods).

The refined RMS displacement from $Acaa$, the root-mean-square shift of all 48 atomic mean positions from the reference, decreases from 0.290 to 0.228~\AA{} on heating and recovers on cooling (Fig.~\ref{fig:atomic-trajectory}c). This path is closely reproduced from different starting points. Framewise refinement initialized from the external neutron structures gives a similar overall path within the same displacement basis. Five of six additional starts for the joint fit, including one close to $Acaa$, also reach nearly the same trajectory, and the sixth stops in a markedly worse minimum (Supplementary Note~4). The heating decrease also persists when the complete doublet and its ratio are omitted from fitting. The displacement remains nonzero at the highest temperature, describing an approach toward $Acaa$ rather than complete recovery of that structure.

We then generated structures at 301 unfitted temperatures and calculated their 213/117 intensity ratios (Fig.~\ref{fig:atomic-trajectory}d). The calculated ratios follow the measured values on both branches, and the measured ratio on cooling lies slightly below that on heating above about 700~K. The ratio is not monotonic along the fitted displacement direction, so its proximity to the $Acaa$ value alone does not measure proximity to that structure. Across the tested fitting conditions, the collective displacement and broad tilt trend are more reproducible than the detailed rotation trajectory. Supplementary Note~4 relates this distinction to the remaining profile residuals and the omitted-reflection checks.

\section{Discussion}

In this work, we developed diffPXRD, a differentiable PXRD engine that includes instrumental and sample effects and evaluates batches of patterns on CPUs or GPUs. Its calculated profiles and derivatives agree with GSAS-II and finite differences. The same engine can refine each pattern independently or fit a shared model to an entire series, and batched evaluation speeds up both. Automatic differentiation also makes the established approach of parametric refinement \cite{stinton2007parametric} more flexible, because a shared model can be changed without writing new derivative code. In addition, because diffPXRD is written in PyTorch, the diffraction calculation can be combined with other differentiable models, such as the reaction and transport models discussed below. Researchers can therefore choose a model that suits the scientific question while keeping the same description of the diffraction experiment.

Batching and parameter sharing provide different benefits. Batching reduces the computing time, and the size of this reduction depends on the workload and hardware, as the benchmarks in Fig.~\ref{fig:performance} and the NCA timings show. Parameter sharing instead allows several measurements to constrain the same coefficients, which makes the refined trajectories more stable when the signal is weak. Because batching alone provides the speedup, joint refinement is not always necessary. Batched independent refinement is useful when each pattern should be interpreted separately, whereas joint refinement is appropriate when the shared relationship is itself of interest or provides a justified constraint on weak signals.

The three applications use different kinds of shared model. In the NCA series, an empirical curve imposes continuity without assuming a microscopic mechanism. In the Li$_2$S series, a kinetic shape adds assumptions about when and how the phase forms and disappears. In Ca$_3$Ti$_2$O$_7$, a displacement basis taken from independent neutron refinements limits the allowed combinations of atomic motion, which reduces the number of unknowns. Each choice determines what can be inferred and how the result should be tested. Reduced variability at low counts shows that a model is stable, whereas predictions at unfitted cycles, reflections, or temperatures test whether the assumed relationship holds beyond the data used to fit it.

A stronger constraint can stabilize a weak signal, but it can also increase the prediction error if it excludes behavior present in the measurements. Model comparison should therefore hold out data for validation and examine both variability and predictive agreement, rather than treating a smoother trajectory as sufficient evidence of a better model. This is particularly important when different structural changes produce similar powder patterns, so that the refinement objective has multiple minima \cite{segal2026loss}. Correlated parameters pose a related problem. A lower site occupancy and a larger displacement parameter, for example, both weaken a site's contribution, but displacement attenuates it more at larger scattering vectors. Automatic differentiation can also supply the residual Jacobian for local estimates of such correlations, and a shared model can reduce them \cite{stinton2007parametric}, for example with one occupancy shared across a temperature series. A physically motivated model can reduce such ambiguity by relating observations across conditions, but its conclusions remain limited by the structural degrees of freedom and experimental effects it includes.

The current implementation is limited to constant-wavelength X-ray diffraction with symmetric pseudo-Voigt, TCH, and FCJ-corrected profiles. Site occupancies, anisotropic displacement parameters, and size and microstrain broadening are implemented and checked (Supplementary Note~1) but not refined here. Time-of-flight, magnetic diffraction, stacking-fault models, and grazing-incidence geometry are not currently supported. Reflection identities and the space group that generates them are held fixed during differentiation. A continuous distortion can be refined in a lower-symmetry setting with separately broadened reflections, whereas a first-order transition with phase coexistence requires a second phase with a refined fraction. For this reason, the Ca$_3$Ti$_2$O$_7$ refinement stops below the first-order transition near 1100~K, where two peaks no longer describe the 213/117 doublet (Supplementary Fig.~S1).

Within this scope, other differentiable models of structural change can supply parameters to the same engine. For example, a numerically integrated reaction or transport model could supply the lattice parameters or phase amounts at each experimental state. A model that predicts the next state from the current one, such as a discretized rate law, could be fitted in the same way, as a sequential counterpart to the joint refinements shown here. Such analyses will become more important as fourth-generation synchrotron sources, such as the Advanced Photon Source Upgrade, increase the rate at which \emph{operando} and \emph{in-situ} series are collected \cite{fornek2019apsu}.

\section{Methods}

\subsection{Forward model and differentiation}

We represent lattice parameters, atomic coordinates, and instrument and sample parameters as PyTorch tensors. Atomic coordinates and tabulated X-ray scattering factors, including anomalous terms, determine the structure factors. We apply multiplicity, Lorentz-polarization, and preferred-orientation corrections to obtain the reflection intensities, then distribute them over the supplied $2\theta$ grid using a peak shape that accounts for instrumental and sample broadening.

The implementation supports three peak-shape options, beginning with a symmetric pseudo-Voigt that combines Gaussian and Lorentzian functions using Caglioti widths and a fixed Gaussian fraction \cite{caglioti1958choice}. The Thompson-Cox-Hastings (TCH) model instead determines the width and mixing parameter from the Gaussian and Lorentzian broadening contributions \cite{thompson1987rietveld}, with the optional FCJ correction adding axial-divergence asymmetry through numerical quadrature.

For each input structure, we determine the intensity scale and background by linear least squares. This procedure, known as variable projection \cite{golub1973variable,oleary2013variable}, gives the following objective for a measured profile $y$ with weight matrix $W$:
\begin{equation}
\ell[y,F(\mathbf{p})]=\min_{a,\beta}\left\|W^{1/2}\left[y-aF(\mathbf{p})-B\beta\right]\right\|_2^2,
\end{equation}
where $\mathbf{p}$ collects the structural and sample parameters, $a$ is the intensity scale, and $B\beta$ is a low-order polynomial background with coefficients $\beta$. Scale and background coefficients are fitted independently for each pattern unless the shared model explicitly links them. In the Li$_2$S analysis, one signal amplitude is shared across each cycle, while the backgrounds remain independent for each pattern.

PyTorch automatic differentiation propagates gradients through the continuous parameters and the linear least-squares solution. Reflection identities and crystallographic symmetry remain fixed during differentiation.

\subsection{Numerical validation}

We select 200 ICSD structures for testing and balance the seven crystal systems and cell-size ranges. The resulting collection contains 149 space groups and 4-192 sites per cell. Structures undergo conventional-cell standardization with pymatgen \cite{ong2013pymatgen} before conversion to the benchmark setting. The numerical validation evaluates the ICSD targets under nine configurations against GSAS-II. Both programs use float64 inputs and anisotropic sample parameters consistent with the crystal symmetry. To compare reflection counts consistently, we use the primary-wavelength angular window while retaining all wavelength components in the profiles. Imported site multiplicities are checked against the symmetry operations in the crystallographic information file (CIF) and normalized where needed.

For ellipsoidal size broadening, both calculations evaluate the size tensor as a quadratic form along the unit scattering vector, and for trigonal $\bar{3}1m$ microstrain both use the published $S_{211}$ basis \cite{vondreele2019broadening}. Supplementary Note~1 specifies the acceptance criteria and derivative tests.

The CuCr$_2$O$_4$ comparison uses 17 measured patterns between 6.8 and 288.7~K from the GSAS-II sequential-refinement tutorial. The primary structure is orthorhombic $Fddd$, with oxygen on the 32h site, and CuO is included as a secondary phase. Each pattern has three independent oxygen coordinates and nine linear coefficients: two phase amplitudes, six Chebyshev background coefficients, and one broad-background-peak amplitude. The instrument parameters are read from the GSAS-II project and verified to match in both programs. Calibrated per-pattern lattice parameters, sample broadening, sample displacement, and broad-background-peak position and width remain fixed. Atomic displacement parameters and starting oxygen coordinates are taken from the first pattern.

GSAS-II uses its native analytic-Jacobian least-squares optimizer. diffPXRD fits the coordinates with the limited-memory Broyden-Fletcher-Goldfarb-Shanno (L-BFGS) optimizer and determines the linear coefficients by variable projection. Serial refinements carry the oxygen coordinates forward, whereas batched refinements start all patterns from the same first-pattern model, and both converge for every pattern. The weighted profile R-factor $R_{\mathrm{wp}}$ uses the supplied point uncertainties. Supplementary Note~1 summarizes the cross-program fit agreement.

\subsection{Computational timings}

The timing benchmarks use the 200 ICSD target structures from the collection described above and compare identical double-precision calculations on four threads of an AMD EPYC 7J13 CPU and one 80~GB A100 GPU in the same node. Complete pattern calculation timings include CIF parsing, reflection setup, device transfer, and profile calculation, with one warm-up and five timed repetitions per structure. A separate timing on the same hardware divides this calculation into stages to measure the setup time. Repeated-evaluation timings reuse the reflection setup and transferred inputs, and they merge symmetry-equivalent reflections through a fixed map built once per structure. Merging is exact while the lattice and the peak-shape model keep the symmetry of the structure, and at representative lattice changes the merged and separately broadened profiles, losses, and lattice gradients agree to within one part in $10^{9}$. Gradients are taken with respect to two lattice-scale parameters per pattern, one for $a$ and $b$ together and one for $c$. Both scales take equal values for cubic cells, so every evaluated lattice keeps the symmetry of its structure. Merged evaluations use batch sizes of 1, 4, 16, and 32, and separately broadened evaluations use batch sizes of 1 and 4. To measure the steady cost within a running refinement, each structure is timed in its own process with the C library set to retain freed memory. Every setting receives four untimed evaluations before five timed repetitions. CPU times come from processes that use only the CPU. A100 times come from processes that evaluate both devices on identical inputs and check that their profiles, losses, and gradients agree outside the timed calculation. Supplementary Note~2 gives the grids, process settings, numerical checks, and timing summaries.

\subsection{NCA lattice model and noise tests}

The NCA analysis uses all 239 patterns in the 18.1-19.8 degree window containing the 003 reflection \cite{mikheenkova2022resolvingdata}. For pattern $s$, measured at capacity $Q_s$ on branch $b_s$, the shared model is
\begin{equation}
\begin{aligned}
z_s &= \sum_{k=1}^{12}\phi_{b_s,k}B_k\!\left(u_{b_s}(Q_s)\right),\\
c_s &= g_\phi(Q_s,b_s)=c_0\exp\!\left[\Delta\tanh\!\left(z_s/\Delta\right)\right].
\end{aligned}
\label{eq:nca_shared_model}
\end{equation}
Here, $b_s$ is either charge or discharge, $u_b(Q)$ rescales the capacity of branch $b$ to the range 0 to 1, and $\phi_{b,k}$ are the fitted coefficients. The cubic B-spline basis functions $B_k$ have equally spaced interior knots and four repeated knots at each end. The hyperbolic tangent keeps $c_s$ within about 5\% of the reference value $c_0=14.2$~\AA{}, since the bound $\Delta=0.05$ gives $c_0e^{-\Delta}<c_s<c_0e^\Delta$. The number of coefficients per branch was chosen from 6, 8, 10, and 12 using only the 191 training profiles, which omit every fifth pattern of each branch. This choice preceded any evaluation of the 48 omitted patterns (Supplementary Note~3) or the noisy realizations. We used the smallest number for which the joint fit stayed within predefined limits relative to framewise fits. These limits were a mean profile loss no more than 20\% higher and $c$ values within 0.005~\AA{} RMS and 0.02~\AA{} at most. Only 12 coefficients met these limits, with a loss ratio of 1.03 and an RMS difference in $c$ of 0.0018~\AA{}. Both methods fit a positive Gaussian and Lorentzian width of the TCH pseudo-Voigt peak, with a fixed FCJ asymmetry, for every pattern. All fits use unweighted least squares, with scale and linear background determined by variable projection. L-BFGS uses a strong-Wolfe line search and at most 200 iterations. The noise comparison uses one independent optimizer per pattern, initialized from its observed peak, without borrowing the joint solution.

Low photon count noise tests are generated by binomial thinning. For each angular bin with integer intensity $I$, we draw $I'\sim\mathrm{Binomial}(I,0.10)$. Original and thinned patterns are rescaled so that their 2nd and 99.5th percentiles map to 0 and 1, and negative values are then set to zero. Joint and framewise refinements use the same twenty realizations, including every pattern without exclusion. At each pattern, we calculate the population standard deviation of the refined $c$ values across these repeats. Trajectory variability is the RMS of these standard deviations over the series. To evaluate agreement with the original profiles, the noisy-fit lattice parameters and widths are held fixed while scale and background are refitted. 

\subsection{Lithium-sulfur intensity model}

The lithium-sulfur analysis uses 131 patterns. We construct the fixed diffraction template using cycles 1-4, determining the antifluorite lattice parameter on a 5.60-5.95~\AA{} grid from the 200 and 111 reflections. The TCH width is set from the published 111 widths, and a fixed intensity scale relates the two reflections.

For normalized half-cycle capacity $u$, precipitation follows $[(u-u_0)/(1-u_0)]_+^{n_g}$, where $[z]_+=\max(z,0)$. The onset $u_0$ sets when precipitation begins, and the exponent $n_g$ determines the subsequent growth curve. Dissolution follows a Johnson-Mehl-Avrami-Kolmogorov type exponential \cite{avrami1939kinetics}, normalized to decrease from one to zero during charge,
\begin{equation}
D(u)=\frac{\exp[-(k u_\epsilon)^{n_d}]-\exp(-k^{n_d})}{\exp[-(k\epsilon)^{n_d}]-\exp(-k^{n_d})},
\label{eq:dissolution}
\end{equation}
with $u_\epsilon=\epsilon+(1-\epsilon)u$ and $\epsilon=10^{-6}$. The scale $k$ and exponent $n_d$ control the decay shape. All four shape parameters are dimensionless.

L-BFGS fits the four shared shape parameters to the 200 reflection in cycles 1-4. These parameters are fixed for cycles 5-8, where one amplitude per cycle is fitted using 200. Linear backgrounds are determined separately for each pattern by variable projection. Evaluation uses the 111 profiles and published integrated areas from cycles 5-8. The monotone spline and framewise amplitudes use the same diffraction template and background treatment. All models are refitted under this protocol for twenty paired binomial realizations at 10\% retention.

\subsection{Atomic motion model}

The atomic motion model for Ca$_3$Ti$_2$O$_7$ expresses the position of atom $j$ as a deviation from the corresponding $Acaa$ reference position,
\begin{equation}
\mathbf{r}_j(T)=\mathbf{r}_{j,Acaa}(T)+\sum_{m=1}^{3}a_m(T)\mathbf{b}_{mj},
\label{eq:atomicbasis}
\end{equation}
where $\mathbf{b}_{mj}$ gives the relative displacement of atom $j$ in vector $m$ and $a_m(T)$ is its amplitude in units of length. The reference and displacement vectors are constructed from a neutron refinement study \cite{kong2023scaling}. Atoms are mapped onto the $A2_1am$ cell before their displacements from $Acaa$ are calculated.

We use singular value decomposition (SVD) to identify the dominant combinations of atomic shifts. The three leading vectors describe nearly all of the squared displacement in the supplied neutron structures and are normalized to unit RMS over the 48 sites. Each amplitude follows a seven-coefficient cubic B-spline with separate heating and cooling branches and a total amplitude bound of 0.4~\AA{}. Supplementary Note~4 gives the spline basis and bound transformation.

To describe the octahedral motions, we align each reference TiO$_6$ octahedron with its refined counterpart by least squares. The rotation components along $c$ are projected onto the neutron-derived rotation pattern and normalized over the octahedra. The in-plane components define the tilt. Displayed structures, structural descriptors, and calculated intensity ratios all use the same refined amplitudes.

\subsection{Atomic motion fitting and evaluation}

The Ca$_3$Ti$_2$O$_7$ experimental archive supplies the lattice parameters and 346 patterns in the selected low-temperature structural interval between 313 and 1069~K \cite{pomiro2020data}. The archive extends to 1223~K, but patterns above 1073~K on heating and above 1055~K on cooling, which include the transition region, are not fitted (Supplementary Note~4). We select 42 patterns nearest to evenly spaced target temperatures, with 21 per branch at about 38~K intervals. This samples the smooth evolution over each branch while reserving the intervening temperatures for interpolation checks. L-BFGS fits the spline coefficients jointly. Framewise fits refine the same amplitudes and per-pattern corrections from external-neutron starts, rather than from the joint solution, against the same observations and objective.

Isotropic atomic displacement parameters are interpolated from the neutron refinements and held at the nearest endpoint outside their temperature range. They describe the spread of an atom around its mean position, distinct from the changes in mean position represented by Eq.~\ref{eq:atomicbasis}. These parameters and the occupancies remain fixed during refinement.

The objective includes weighted least-squares profile residuals in ten reflection windows, using the supplied point variances. We fit one nonnegative scale per pattern and one linear background per window. To increase the weight of the relative doublet intensities, we add the squared difference between calculated and measured integrated-intensity ratios, divided by the squared ratio uncertainty. This uncertainty comes from the two-peak fit, with a relative floor of 2\%. The ratio term reweights information from the fitted windows rather than adding an independent measurement. Supplementary Note~4 specifies those windows, ratio extraction, reconstruction of predictions, and sensitivity tests.

\section*{Data availability}
The measured series are available at \url{https://doi.org/10.5281/zenodo.7049855}, \url{https://doi.org/10.5281/zenodo.3514967}, and \url{https://doi.org/10.6084/m9.figshare.12446306.v1}. The CuCr$_2$O$_4$ patterns are distributed with the official GSAS-II sequential-refinement tutorial, and the Ca$_3$Ti$_2$O$_7$ neutron structures are taken from the supplementary tables of ref.~\citenum{kong2023scaling}. Figure source data, numerical benchmarks, and exact analysis configurations accompany the code release.

\section*{Code availability}
The diffPXRD implementation and analysis scripts are available at \url{https://github.com/AdvancedPhotonSource/DiffPXRD}. 

\section*{Acknowledgements}
This research used resources of the Advanced Photon Source, a U.S.~Department of Energy (DOE) Office of Science user facility at Argonne National Laboratory, and is based on research supported by the U.S. DOE Office of Science-Basic Energy Sciences, under Contract No.~DE-AC02-06CH11357.

\section*{Competing interests}
The authors declare no competing interests.
\bibliographystyle{unsrtnat}
\bibliography{references}
\clearpage
\includepdf[pages=-,fitpaper=true]{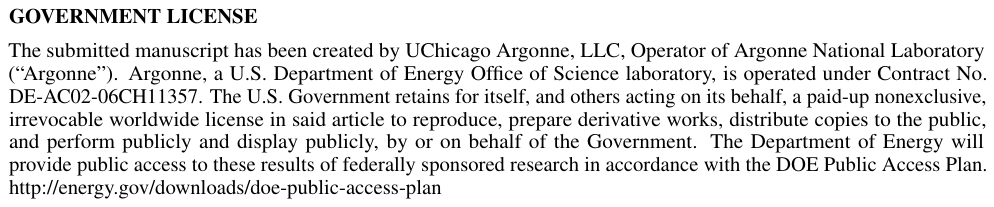}
\end{document}